\documentclass[12pt,twocolumn]{iopart}
\usepackage{graphicx,bbm,amssymb}

\newcommand{\be}{\begin{equation}}
\newcommand{\ee}{\end{equation}}

\newcommand{\bra}[1]{{\langle #1 \vert}}

\newcommand{\ket}[1]{{\vert #1 \rangle}}

\newcommand{\ave}[1]{{\langle #1\rangle}}
\newcommand{\ii}{ {\rm i} }
\newcommand{\dd}{ {\rm d} }

\newcommand{\RaR}{\mathbb{R}}
\newcommand{\CC}{\mathbb{C}}

\newcommand{\z}{{\rm z}}

\newcommand{\mm}[1]{{\mathbf{#1}}}

\newcommand{\LL}{{\hat{\cal L}}}
\newcommand{\DD}{{\hat{\cal D}}}
\newcommand{\VV}{{\hat {\cal V}}}

\def\tr{{{\rm tr}}}
\def\ad{{\,{\rm ad}\,}}
\def\one{\mathbbm{1}}

\newcommand{\La}{{\mathtt L}}
\newcommand{\Ra}{{\mathtt R}}

\begin{document}

\title[Comments on boundary driven open XXZ chain]{Comments on boundary driven open XXZ chain: asymmetric driving and uniqueness of steady states}
\author{Toma\v{z} Prosen}
\address{Department of Physics, FMF,  University of Ljubljana, Jadranska 19, 1000 Ljubljana, Slovenia}
\eads{\mailto{tomaz.prosen@fmf.uni-lj.si}}

\date{\today}

\begin{abstract}
In this short note we provide two extensions on the recent explicit results on the matrix-product ansatz for the non-equilibrium steady state of a markovianly boundary-driven anisotropic Heisenberg XXZ spin 1/2 chain. We write a perturbative solution for the steady state density matrix in the system-bath coupling for an arbitrary (asymmetric) set of four spin-flip rates at the two chain ends, generalizing the symmetric-driving ansatz of [Phys.~Rev.~Lett.~{\bf 106},~217206~(2011)].  Furthermore, we generalize the exact (non-perturbative) form of the steady state for just two Lindblad channels (spin-up flipping on the left, and spin-down flipping on the right) to an arbitrary (asymmetric) ratio of the spin flipping rates [Phys.~Rev.~Lett.~{\bf 107},~137201~(2011)]. In addition, we also indicate a simple proof of uniqueness of our steady states.
\end{abstract}

 
%\maketitle

\section{Introduction}

The anisotropic Heisenberg (XXZ) model \cite{heisenberg} of $n$ coupled quantum spins 1/2 with the Hamiltonian
\begin{equation}
H = \sum_{j=1}^{n-1} \left( 2\sigma^+_j \sigma^-_{j+1} + 2\sigma^-_j \sigma^+_{j+1} + \Delta \sigma^{\rm z}_j \sigma^{\rm z}_{j+1} \right)
\label{eq:hamXXZ}
\end{equation}
can be considered as a prototype of a many body quantum model of strong interactions. We write Pauli operators on a tensor product space ${\cal F}_n=(\CC^2)^{\otimes n}$, as
$\sigma^s_j = \one_{2^{j-1}}\otimes \sigma^s \otimes \one_{2^{n-j}}$, $\one_d$ being a $d$-dimensional unit matrix, where
$\sigma^\pm = \frac{1}{2}(\sigma^{\rm x} \pm\ii \sigma^{\rm y})$ and $\sigma^{\rm x,y,z}$ are the standard $2\times 2$ Pauli matrices.

The XXZ model exhibits a rich variety of equilibrium and nonequilibrium physical behaviors. In nature it provides an excellent description of the so called spin-chain materials \cite{spinchain}, and it is believed to provide the key for understanding of various collective quantum phenomena in low dimensional strongly interacting systems, such as magnetic or superconducting transitions in two dimensions. Although equilibrium (thermodynamic) properties of XXZ chain are well understood in terms of Bethe Ansatz \cite{takahashi}, as the model represents a paradigmatic example of quantum integrable systems, its nonequilibrium properties at finite temperature are lively debated \cite{affleck}. 

We consider the non-equilibrium quantum transport model \cite{wichterich} based on a Markovian master equation for the XXZ chain in the Lindblad form \cite{lindblad,breuer}
\be
\frac{\dd \rho(t)}{\dd t} = -\ii [H,\rho(t)] + \sum_k 2 L_k \rho(t) L^\dagger_k - \{ L^\dagger_k L_k,\rho(t)\}
\label{eq:lindblad}
\ee
with a set of four boundary-supported Lindblad operators
\begin{equation}
L_1 = \sqrt{\alpha} \sigma^-_1, \quad L_2 = \sqrt{\beta} \sigma^+_1, \quad
L_3 = \sqrt{\gamma} \sigma^-_n, \quad L_4 = \sqrt{\delta} \sigma^+_n
\label{eq:lop}
\end{equation}
where $\alpha,\beta,\gamma,\delta$, represent, respectively, spin-down/spin-up incoherent transition rates at the left and right boundary of the chain. These are assumed to be the only incoherent processes in the model, whereas its bulk dynamics is fully specified by the Hermitian many-body Hamiltonian $H$.

In particular we are interested in the {\em non-equilibrium steady state} (NESS), with the density operator $\rho_\infty = \lim_{t\to\infty} \rho(t)$ satisfying the fixed point condition:
\begin{equation}
\LL \rho_\infty = 0, 
\label{eq:fixedpoint}
\end{equation}
where $\LL$ is the Liouvillean, decomposed into the unitary, and dissipative parts:
\begin{eqnarray}
&& \LL := -\ii \ad H + \DD,  \label{eq:fixed} \\
&& \DD := \alpha \DD_{\sigma^-_1} + \beta \DD_{\sigma^+_1} +  \gamma \DD_{\sigma^-_n} + \delta \DD_{\sigma^+_n},\\
&& \DD_\sigma (\rho) := 2 \sigma \rho \sigma^\dagger - \{\sigma^\dagger \sigma,\rho\}.
\end{eqnarray}

The boundary driven open XXZ model (\ref{eq:lindblad}) with (\ref{eq:hamXXZ},\ref{eq:lop}) can be considered as a hybrid classical/quantum Markovian lattice gas 
(see e.g. \cite{eisler} for related ideas), namely the boundary injection/absorbtion rates are the same as in some classical Markovian stochastic many-body processes (i.e. simple exclusion processes \cite{blythe,schutz}) while the bulk dynamics is fully coherent. Note that the effect of incoherent processes on the boundary, which works against developing strong macroscopic entanglement in the course of time-evolution, also enables efficient applicability of Liouville space density-matrix-renormalization group methods for computing NESSes for generic (non-solvable) local spin chain Hamiltonians \cite{pz}. The applicability of incoherent boundary processes to model the (magnetic) baths can be indeed justified if there is a finite correlation (coherence) length in the microscopic model of the baths.

Below we make a few remarks concerning recently developed exact solution for NESS of the boundary driven XXZ model \cite{new,new2}. In section \ref{uniqueness} we show how uniqueness of NESS simply follows from a theorem of Evans \cite{evans}. In section \ref{pert} we then extend the zeroth and the first order of NESS density matrix in the weak-coupling perturbative expansion \cite{new} to the case of arbitrary boundary spin-flipping rates $\alpha,\beta,\gamma,\delta$, while in section \ref{nonpert} we in a similar way extend the non-perturbative exact solution of Ref.~\cite{new2}.

\section{Proof of uniqueness of the steady state}

\label{uniqueness}

Let us first show that under quite general conditions, the open-XXZ model (\ref{eq:lindblad}) with (\ref{eq:hamXXZ},\ref{eq:lop}), possesses a {\em unique} NESS (\ref{eq:fixed}), i.e. the fixed point $\rho_\infty$ is independent of the initial state $\rho(0)$. 

We start by noting a theorem of Evans \cite{evans} (which is a generalization of Ref.~\cite{spohn}) which essentially states that NESS is unique iff the set of operators $\{H,L_1,L_2,\ldots\}$ generates, under multiplication and addition, the entire algebra of (bounded) operators, in our case the Pauli algebra ${\cal B}({\cal F}_n)$ of the spin-1/2 chain on $n$ sites. Indeed, this is true even if we take only the Hamiltonian $H$, and a single pair of one up-flip and one down-flip Lindblad operators out of four (\ref{eq:lop}), say $\sigma^+_1$, and $\sigma^-_1$. Note that the scalar prefactors $\sqrt{\alpha},\sqrt{\beta}$ are not important for this discussion as we are only interested in the generators of the algebras and not in the operators themselves.

One then observes the following recursive operator identities:
\begin{eqnarray}
\sigma^+_2  &=& \frac{1}{4}\sigma^{\rm z}_1 [\sigma^+_1,[H,\sigma^{\rm z}_1]], \label{eq:opid1}\\
\sigma^+_{j} &=& -\sigma^+_{j-2}  - \frac{1}{2}\sigma^{\rm z}_{j-1} [\sigma^-_{j-1},\sigma^+_{j-1} H \sigma^+_{j-1}], \quad j=3,4\ldots, n,\label{eq:opid2}
\end{eqnarray}
which generate the entire set $\{\sigma^+_j; j=1,\ldots,n\}$ starting from just $H$ and $\sigma^+_1$. Similarly, 
$\{\sigma^-_j; j=1,\ldots,n\}$ are generated by Hermitian adjoints of (\ref{eq:opid1},\ref{eq:opid2}).
Clearly, the set $\{\sigma^+_j,\sigma^-_j; j=1,\ldots,n\}$ then generates, by multiplication and addition, the entire algebra ${\cal B}({\cal F}_n)$.
We note that we can also take only a pair up-flip and down-flip Lindblad operators from the opposite edges, say $\sigma^+_1$ and $\sigma^-_n$, i.e. if $\alpha=\delta=0$ (or, alternatively, if $\beta=\gamma=0$), since the operator recurrence (\ref{eq:opid1},\ref{eq:opid2}), or its Hermitian conjugate, can also be started at $j=n$ (i.e. it is symmetric with respect to the replacement $j \leftrightarrow n+1-j$).

\section{Perturbative (weak coupling) solution}

\label{pert}

We start by considering the weak-coupling regime, where all the four rates are small,
\be
\alpha=\varepsilon  a, \quad \beta=\varepsilon b,\quad \gamma = \varepsilon c,\quad \delta = \varepsilon d
\ee
and $\varepsilon$ is considered as a small parameter.
Let us write the NESS density operator as a formal power series
\be 
\rho_\infty = \sum_{p=0}^\infty (\ii \varepsilon)^p \rho^{(p)}.
\label{eq:pertansatz}
\ee
Plugging the ansatz (\ref{eq:pertansatz}) into the fixed point condition (\ref{eq:fixedpoint})
results in an operator valued recurrence relation for the sequence $\{ \rho^{(p)};p=0,1,2\ldots \}$
\be
[H,\rho^{(p)}] = 
\left\{ \begin{array}{ll} 0, & {\rm if\;} p=0; \cr
-\DD_0 \rho^{(p-1)}, & {\rm if\;} p=1,2,\ldots \end{array} \right.
\label{eq:recurrence}
\ee
where $\DD_0 = a \DD_{\sigma^-_1} + b \DD_{\sigma^+_1} +  c \DD_{\sigma^-_n} + d \DD_{\sigma^+_n}$,
so that $\DD \equiv \varepsilon \DD_0$.
In Ref. \cite{new} we have shown that for a particular case of symmetric driving 
\begin{equation}
a_{\rm sym}\equiv d_{\rm sym} \equiv \frac{1-\mu}{2},\qquad
b_{\rm sym}\equiv c_{\rm sym} \equiv \frac{1+\mu}{2},
\label{eq:sym}
\end{equation}
one can express the zeroth, the first, and the second order of the perturbation series
\begin{eqnarray}
2^n \rho^{(0)}_{\rm sym} &=& \one, \label{eq:zorder} \\
2^n \rho^{(1)}_{\rm sym} &=& \mu (Z - Z^\dagger), \label{eq:forder} \\
2^n \rho^{(2)}_{\rm sym} &=& \frac{\mu^2}{2} (Z - Z^\dagger)^2 - \frac{\mu}{2} [Z,Z^\dagger]. \label{eq:sorder}
\end{eqnarray}
in terms of a non-Hermitian {\em matrix product operator}
\be 
Z = \!\!\!\!\!\sum_{(s_1,\ldots,s_n)\in\{+,-,0\}^n}\!\!\!\!\!\bra{\La} \mm{A}_{s_1}\mm{A}_{s_2}\cdots\mm{A}_{s_n}\ket{\Ra} 
\sigma^{s_1}\otimes\sigma^{s_2} \cdots\otimes \sigma^{s_n},
\label{eq:MPZ}
\ee
where $\sigma^{0}_j \equiv \one$. $\mm{A}_0, \mm{A}_\pm$ is a triple of near-diagonal matrix operators acting on an auxiliary Hilbert space
${\cal H}$ spanned by an orthonormal basis $\{ \ket{\La},\ket{\Ra},\ket{1},\ket{2},\ldots \}$:
\begin{eqnarray}
\mm{A}_0 &=& 
\ket{\La}\bra{\La} + \ket{\Ra}\bra{\Ra} + \sum_{r=1}^\infty \cos\left(r\lambda\right) \ket{r}\bra{r}, \nonumber \\
\mm{A}_+ &=& \ket{\La}\bra{1} + c \sum_{r=1}^\infty \sin\left(2\left\lfloor \frac{r\!+\!1}{2}\right\rfloor \lambda\right) \ket{r}\bra{r\!+\!1},\label{eq:explicitA1}\\
 \mm{A}_- &=& \ket{1}\bra{\Ra} - c^{-1}\sum_{r=1}^\infty \sin\left(\!\left(2\left\lfloor \frac{r}{2}\right\rfloor\!+\!1\right)\lambda\right)\ket{r\!+\!1}\bra{r}, \nonumber
\end{eqnarray}
where $\lambda = \arccos \Delta \in \RaR \cup \ii\RaR$ and $\lfloor x \rfloor$ is the largest integer not larger than $x$.
Constant $c\in\CC-\{0\}$ is arbitrary, but it is perhaps suitable to choose $c=1$ for $|\Delta| \le 1$ ($\lambda\in \RaR$) and $c=\ii$ for $|\Delta| > 1$ ($\lambda\in \ii\RaR$) making the matrices (\ref{eq:explicitA1}) {\em always real}.
The key property of the Z operator which is proven and used extensively in \cite{new} is the almost-commutation (or conservation law) property
\be
[H,Z] = -\sigma^{\rm z}_1 + \sigma^{\rm z}_n.
\label{eq:HZ}
\ee

We shall proceed to show now that this solution can be used also to express (perturbatively) the leading orders of NESS density operator for arbitrary drivings (arbitrary rates $a,b,c,d$). This we will do by writing quite general ans\" atze for the zeroth and the first order
\begin{eqnarray}
2^n \rho^{(0)} &=& (\sigma^0 + \nu \sigma^{\rm z})^{\otimes n} =: R, \label{eq:zorder1}\\
2^n \rho^{(1)} &=& \mu (Z - Z^\dagger) R \label{eq:forder1},
\end{eqnarray}
where $\nu$ (related to an average magnetization in the zeroth order $\nu=\ave{\sigma^{\rm z}_j}_{\varepsilon\to 0}$) and $\mu$ are still undetermined parameters. We note that the operator $R$ can be expressed in terms of an exponentiated total spin projection, 
\be
R=\sqrt{1-\nu^2}\exp\{({\rm artanh}\,\nu) M^\z\},\qquad M^\z = \sum_{j=1}^n\sigma^{\rm z}_j,
\ee
hence it can be shown to commute with the Hamiltonian (\ref{eq:hamXXZ}) and the Z-operator (\ref{eq:MPZ}) \footnote{$[Z,M^\z]=0$ follows from the fact that all the Pauli terms of (\ref{eq:MPZ}) contain the same number of $\sigma^+$ and $\sigma^-$ tensor-factors.}
\begin{equation}
[H,R]=0,\quad [Z,R]=0, 
\label{eq:conditions2}
\end{equation}
guaranteeing the zerorth order $p=0$ condition (\ref{eq:recurrence}) and
making the ordering of the terms in the first order (\ref{eq:forder1}) not important.

Plugging the ans\" atze (\ref{eq:zorder1},\ref{eq:forder1}) into the equation $[H,\rho^{(1)}]=-\DD \rho^{(0)}$ and using the equations (\ref{eq:HZ},\ref{eq:conditions2}) one gets an identitiy
$\mu \sigma^{\rm z}_1 - \mu \sigma^{\rm z}_n = \frac{1}{1-\nu^2}\{(b-a-\nu(a+b))\sigma^{\rm z}_1+(d-c-\nu(c+d))\sigma^{\rm z}_n-
\nu(b-a-\nu(a+b)+d-c-\nu(c+d))\one\}$, which immediately fixes the unknown parameters $\nu,\mu$.
Namely, the {\em average magnetization} reads
\be
\nu=\frac{b+d-a-c}{a+b+c+d},
\ee 
and the {\em effective driving} is
\be
\mu=\frac{2(bc-ad)}{(1-\nu^2)(a+b+c+d)}.
\ee
We note that the ansatz for the second order $\rho^{(2)}$ (\ref{eq:sorder}) cannot be extended to general asymmetric boundary conditions in a similar way as the zeroth and the first orders $\rho^{(p)} = R \rho^{(p)}_{\rm sym}$,
$p\in\{0,1\}$.

\section{Non-perturbative (extreme driving) solution}

\label{nonpert}

Let us now focus on a nonperturbative (exact) solution of the case with just two Lindblad channels
\be
L_1 = \sqrt{\beta} \sigma^+_1,\quad L_2 = \sqrt{\gamma} \sigma^-_n
\label{eq:lop2}
\ee
For the symmetric situation $\beta=\gamma=:\varepsilon$ this corresponds to the extreme driving case $\mu=1$ of (\ref{eq:sym}) which has been solved in exactly in Ref.~\cite{new2}, namely
\be
\rho^{\rm sym}_\infty = \frac{S_n S^\dagger_n}{\tr ( S_n S^\dagger_n )}
\label{eq:defR}
\ee
and $S_n$ is a non-Hermitian {\em matrix product operator}
\be 
S_n = \!\!\!\!\!\!\sum_{(s_1,\ldots,s_n)\in\{+,-,0\}^n}\!\!\!\!\!\bra{0} \mm{A}'_{s_1}\mm{A}'_{s_2}\cdots\mm{A}'_{s_n}\ket{0} \sigma^{s_1}\otimes \sigma^{s_2} \cdots\otimes \sigma^{s_n}
\label{eq:MPS}
\ee
where $\sigma^{0} \equiv \one_2$ and $\mm{A}'_0, \mm{A}'_\pm$ is a triple of near-diagonal matrix operators acting on an infinite-dimensional auxiliary Hilbert space
${\cal H}'$ spanned by an orthonormal basis $\{ \ket{0},\ket{1},\ket{2},\ldots \}$:
\begin{eqnarray}
\mm{A}'_0 &=& \ket{0}\bra{0} + \sum_{r=1}^\infty a^0_r \ket{r}\bra{r}, \nonumber \\
\mm{A}'_+ &=& \ii \varepsilon \ket{0}\bra{1} + \sum_{r=1}^\infty a^+_r \ket{r}\bra{r\!+\!1},\label{eq:explicitA}\\
 \mm{A}'_- &=& \ket{1}\bra{0} + \sum_{r=1}^\infty a^-_r \ket{r\!+\!1}\bra{r}, \nonumber
 \end{eqnarray}
with matrix elements (writing again $\lambda = \arccos \Delta$)
\begin{eqnarray}
a^0_r        &=& \cos\left(r\lambda\right) + \ii\varepsilon \frac{ \sin\left(r \lambda \right)}{2\sin \lambda}, \nonumber\\
a^+_{2k-1} &=& c\sin\left(2k\lambda\right)+ \ii\varepsilon \frac{c \sin\left((2k\!-\!1)\lambda\right)\sin\left(2k\lambda\right)}{2(\cos\left((2k\!-\!1)\lambda\right)+\tau_{2k-1})\sin\lambda},  \nonumber \\
a^+_{2k}     &=&  c\sin\left(2k\lambda\right)- \ii\varepsilon \frac{c (\cos\left(2k\lambda\right)+\tau_{2k})}{2\sin\lambda}, \label{eq:a} \\
a^-_{2k-1}  &=& -\frac{\sin\left((2k\!-\!1)\lambda\right)}{c} + \ii\varepsilon \frac{\cos\left((2k\!-\!1)\lambda\right)+\tau_{2k-1}}{2c\sin\lambda}, \nonumber \\
a^-_{2k}      &=& -\frac{\sin\left((2k\!+\!1)\lambda\right)}{c} - \ii\varepsilon \frac{ \sin\left(2k\lambda\right)\sin\left((2k\!+\!1)\lambda\right)}{2c(\cos\left(2k\lambda\right)+\tau_{2k})\sin\lambda}. \nonumber 
\end{eqnarray}
Constant $c \in \CC - \{0\}$ and signs $\tau_r \in \{\pm 1\}$ are arbitrary,
i.e. all choices of $c,\tau_r$, for $r=1,2,\ldots$, give identical operator $S_n$ (\ref{eq:MPS}).

The crucial ingredient in the proof of Ref.~\cite{new2} was the following recursive identity satisfied by operators $S_n$
\be
[H,S_n] = -\ii \varepsilon (\sigma^{\rm z}\otimes S_{n-1} - S_{n-1}\otimes \sigma^{\rm z}),
\label{eq:HS}
\ee
which can be understood as a `non-perturbative' analog of the commutator relation (\ref{eq:HZ}).

We shall now show how one can generalize the ansatz (\ref{eq:defR},\ref{eq:MPS}) in order to incorporate the asymmetric driving (\ref{eq:lop2}) for any rates $\beta,\gamma$. We start by recognizing the following non-unitary symmetry of the XXZ dynamics:\\\\
{\bf Lemma:} 
{\em
Let $\nu\in(-1,1)$ be a real parameter and $\VV_1 : {\cal B}({\cal F}_1) \to {\cal B}({\cal F}_1)$ a non-unitary, but non-degenerate linear map of a set of $2\times 2$ matrices onto itself, which is 
completely specified by its action on the Pauli basis $\tilde{\sigma}^s = \VV_1(\sigma^s)$, $s\in\{ 0,+,-,{\rm z}\}$:
\begin{eqnarray}
\tilde{\sigma}^\pm &=& \sigma^\pm,\nonumber \\ 
\tilde{\sigma}^0 &=&  \frac{1}{\sqrt{1-\nu^2}}( \sigma^0 - \nu \sigma^{\rm z}), \label{eq:transf}\\
\tilde{\sigma}^{\rm z} &=& \frac{1}{\sqrt{1-\nu^2}}(\sigma^{\rm z}-\nu \sigma^0).\nonumber
\end{eqnarray}
$\VV_1$ induces a one-to-one linear map $\VV = \VV_1^{\otimes n}$ of $n$-spin Pauli algebra ${\cal B}({\cal F}_n)$, 
which is completely specified by $\VV(\sigma^{s_1}\otimes \sigma^{s_2}\cdots\otimes \sigma^{s_n}) = 
\tilde{\sigma}^{s_1}\otimes \tilde{\sigma}^{s_2}\cdots\otimes\tilde{\sigma}^{s_n}$.
Then, $\VV$ commutes with the Heisenberg dynamics of the XXZ chain (\ref{eq:hamXXZ}), i.e. for any $x\in {\cal B}({\cal F}_n)$:
\be
[H,\VV(x)] = \VV([H,x]).
\label{eq:lemma}
\ee
}
\noindent {\bf Proof:}
Writing the Hamiltonian (\ref{eq:hamXXZ}) as a sum of two body terms $H=\sum_{j=1}^{n-1}h_{[j,j+1]}$, a sufficient condition for (\ref{eq:lemma}) to
hold is $[h_{[j,j+1]},\VV(x)]= \VV([h_{[j,j+1]},x])$. Due to the linearity of this relation in $x$, the latter can be considered to be of the form
$x=x_{[1,j-1]}\otimes x_{[j,j+1]} \otimes x_{[j+2,n]}$, where the first, or the last, factor are taken as trivial if $j=1$, or $j=n-1$, respectively.
Proving the lemma is then equivalent to showing (\ref{eq:lemma}) for $n=2$, i.e. 
\be
[h,\tilde{\sigma}^s \otimes \tilde{\sigma}^t] = \VV([h,\sigma^s\otimes\sigma^t]), \quad s,t\in\{0,+,-,\z\}
\ee
where $h = 2 \sigma^+ \otimes \sigma^- +  2 \sigma^- \otimes \sigma^+ + \Delta \sigma^\z \otimes \sigma^\z$.
This follows from observing:
\begin{eqnarray}
[h,\sigma^+\otimes\sigma^0] &=& 2\Delta \sigma^+ \otimes \sigma^\z - 2\sigma^\z\otimes \sigma^+, \cr
[h,\sigma^+\otimes\sigma^\z] &=& 2\Delta \sigma^+ \otimes \sigma^0 - 2\sigma^0\otimes \sigma^+, \cr
[h,\sigma^+\otimes\sigma^-] &=& \sigma^0 \otimes \sigma^\z - \sigma^\z\otimes \sigma^0, \label{eq:x} \cr
[h,\sigma^0\otimes\sigma^\z] &=& 4\sigma^+ \otimes \sigma^- - 4\sigma^-\otimes \sigma^+,  \cr
[h,\sigma^s\otimes\sigma^s]  &=& 0, \quad {\rm for} \quad s\in\{0,+,-,\z\}, 
\end{eqnarray}
together with related Hermitian conjugate identities and identities with swapped tensor factors, and checking exactly identical identities with $\sigma^s$ replaced by $\tilde{\sigma}^s$. QED

We note that the non-unitary symmetry map $\VV$, which is in fact a peculiar representation of the Lorentz group $SO(1,1)$, is  {\em not} a canonical transformation, namely $\tilde{\sigma}^s$ {\em do not} satisfy the same commutation relations as the Pauli matrices $\sigma^s$.

Now, we are in a position to state the main result:\\\\
{\bf Theorem:} {\em The unique NESS of the flow (\ref{eq:lindblad}) with (\ref{eq:hamXXZ},\ref{eq:lop2}) can be written as
\be
\rho_\infty = \frac{ \tilde{S}_n \tilde{S}_n^\dagger R}{\tr (\tilde{S}_n \tilde{S}_n^\dagger R)}
\label{eq:genansatz}
\ee
where $R = (\sigma^0 + \nu \sigma^{\rm z})^{\otimes n}$ is the (unnormalized) weak-coupling limit (\ref{eq:zorder1}), 
and the operators $\tilde{S}_n$ are given in terms of the same form of the MPO as in the symmetric case (\ref{eq:MPS})
\be 
\tilde{S}_n = \!\!\!\!\!\!\!\!\!\!\!\!\!\sum_{(s_1,\ldots,s_n)\in\{+,-,0\}^n}\!\!\!\!\!\!\!\!\!\!\!\!\!\!\bra{0} \mm{A}'_{s_1}\mm{A}'_{s_2} \cdots\mm{A}'_{s_n} \ket{0} \tilde{\sigma}^{s_1}\otimes \tilde{\sigma}^{s_2} \cdots\otimes \tilde{\sigma}^{s_n}
\label{eq:MPS1}
\ee
but in a different basis (\ref{eq:transf}), i.e. $\tilde{S}_n = \VV(S_n)$, with the magnetization parameter:
\be
\nu = \frac{\beta-\gamma}{\beta+\gamma}, \label{eq:nu}
\ee
 and the effective coupling $\varepsilon$ (to be used in explicit expressions (\ref{eq:explicitA},\ref{eq:a}) for the triple of matrices $\mm{A}'_+,\mm{A}'_-,\mm{A}'_0$):
\be
\varepsilon = \sqrt{\beta \gamma}. \label{eq:eps}
\ee
}

\noindent
{\bf Proof:}
Uniqueness of the fixed point $\rho_\infty$ has been shown, also for the restricted case of driving (\ref{eq:lop2}), in section \ref{uniqueness}.

We need to show that (\ref{eq:genansatz}) satisfies the equation (\ref{eq:fixedpoint}), which, together with $[R,\tilde{S}_n]=0$ again following from the fact that all the Pauli terms of (\ref{eq:MPS1}) contain the same number of 
$\sigma^+ \equiv \tilde{\sigma}^+$ and $\sigma^-\equiv\tilde{\sigma}^-$ tensor-factors, is equivalent to the condition
\be
\ii [H,\tilde{S}_n \tilde{S}_n^\dagger] = \DD(\tilde{S}_n \tilde{S}_n^\dagger R)R^{-1}
\label{eq:key1}
\ee
Note that, since $\tilde{S}_n = \VV(S_n)$, the Lemma implies
\be
[H,\tilde{S}_n]=-\ii \tilde{\varepsilon} \left((\sigma^\z - \nu \sigma^0)\otimes \tilde{S}_{n-1} - \tilde{S}_{n-1}\otimes (\sigma^\z - \nu\sigma^0)\right),
\label{eq:modHS}
\ee
where $\tilde{\varepsilon} = \varepsilon/\sqrt{1-\nu^2}$ and $\varepsilon$ is still unspecified.
Following the idea in the proof of Ref.~\cite{new2}, we
can use the algebraic relations among $\mm{A}'_s$ to write
$\tilde{S}_n =: \tilde{\sigma}^0 \otimes \tilde{S}_{n-1} + \tilde{\sigma}^+ \otimes \tilde{P}_{n-1} =: \tilde{S}_{n-1}\otimes \tilde{\sigma}^0 + \tilde{Q}_{n-1} \otimes \tilde{\sigma}^-$, 
defining $\tilde{Q}_{n-1}$ and $\tilde{P}_{n-1}$ as some linear operators over ${\cal F}_{n-1}$. This merely expresses the fact that in $\tilde{S}_n$ the first 
$\sigma^+$ always comes to the left of all $\sigma^-$, and the last $\sigma^-$ comes to the right of all $\sigma^+$.
Straightforward calculation, employing (\ref{eq:modHS}) and definitions of $\tilde{P}_n,\tilde{Q}_n$, then expresses the left-hand-side of (\ref{eq:key1}) as
\begin{eqnarray}
\!\!\!\!\!\!\!\!\!\!\!\!\!\!\!\!\!\!\!\!\!\!\!\!\!\!\!\!\!\!&&\tilde{\varepsilon}\Bigl\{
2(\sigma^\z - \nu \sigma^0)\otimes\tilde{S}_{n-1}\tilde{S}^\dagger_{n-1} -(\nu+1) \sigma^+ \otimes \tilde{P}_{n-1} \tilde{S}^\dagger_{n-1}
-(\nu+1) \sigma^-\otimes\tilde{S}_{n-1}\tilde{P}^\dagger_{n-1} \nonumber \\
\!\!\!\!\!\!\!\!\!\!\!\!\!\!\!\!\!\!\!\!\!\!\!\!\!\!\!\!\!\!&&-2 \tilde{S}_{n-1} \tilde{S}_{n-1}^\dagger \otimes (\sigma^\z - \nu \sigma^0)
+(\nu-1) \tilde{Q}_{n-1}\tilde{S}^\dagger_{n-1}\otimes \sigma^- +(\nu-1)\tilde{S}_{n-1}\tilde{Q}_{n-1}^\dagger \otimes \sigma^+ \Bigr\}.
\label{eq:lhs}
\end{eqnarray}
For the right-hand-side of (\ref{eq:key1}), we use the fact that the map 
\be
\rho \to \DD(\rho R)R^{-1}
\ee 
nontrivially acts only locally, on sites $j=1$ and $j=n$, finding
\begin{eqnarray}
\!\!\!\!\!\!\!\!\!\!\!\!\!\!\!\!\!&& \frac{2\beta}{1+\nu} (\sigma^\z - \nu \sigma^0) \otimes \tilde{S}_{n-1}\tilde{S}_{n-1}^\dagger - \beta \sigma^+ \otimes \tilde{P}_{n-1} \tilde{S}^\dagger_{n-1} - \beta \sigma^- \otimes \tilde{S}_{n-1}\tilde{P}_{n-1}^\dagger  \nonumber \\
\!\!\!\!\!\!\!\!\!\!\!\!\!\!\!\!\!&-&\frac{2\gamma}{1-\nu} \tilde{S}_{n-1}\tilde{S}^\dagger_{n-1}\otimes (\sigma^\z - \nu\sigma^0) - \gamma \tilde{Q}_{n-1}\tilde{S}^\dagger_{n-1}\otimes\sigma^- - \gamma \tilde{S}_{n-1}\tilde{Q}^\dagger_{n-1}\otimes \sigma^+. 
\label{eq:rhs}
\end{eqnarray}
Indeed, these two expressions, (\ref{eq:lhs}) and (\ref{eq:rhs}), are linear combinations of exactly the same terms.
Comparing term-wise results in conditions $\tilde{\varepsilon} = \beta/(1+\nu) = \gamma/(1-\nu)$, yielding, with $\varepsilon = \sqrt{1-\nu^2}\tilde{\varepsilon}$, exactly the expressions (\ref{eq:nu}) and (\ref{eq:eps}). QED

\section{Conclusion}

Computation of expectation values of physical observables in NESS, based on our new results for generalized boundary conditions, can be facilitated in terms of the transfer matrices which can be constructed exactly as in Refs.~\cite{new,new2}. The effect of the asymmetric boundary conditions is essentially an offset of average magnetization, which is a consequence of the non-unitary symmetry $\VV$ described in the Lemma of section \ref{nonpert}.

Inspiring discussions with W. de Roeck, H. Spohn and F. Verstraete, as well as the hospitality of NORDITA in Stockholm where a part of the results reported here were obtained, are warmly acknowledged. The work has been  supported by the grants J1-2208 and P1-0044 of ARRS (Slovenia), and ESF (supporting NORDITA program).

\end{document}